\documentclass[hyper]{JHEP3}

\input{epsf}
\usepackage{epsfig}
\usepackage{amssymb}
\usepackage{amsfonts}
\usepackage{amsbsy}
\usepackage[all,2cell]{xy}

\usepackage{amsmath}

\usepackage{amssymb,amscd}
\usepackage{mathrsfs}
\usepackage{amsmath,amsthm}
\usepackage{manfnt}

\def\be{\begin{equation}}
\def\ee{\end{equation}}

\def\mod{{\rm mod}}
\def\Tr{{\rm Tr}}

\def\IC{\mathbb{C}}

\def\IQ{\mathbb{Q}}
\def\IR{{\mathbb{R}}}

\def\IZ{{\mathbb{Z}}}

\def\CC {{\cal C}}
\def\CM {{\cal M}}

\def\CN {{\cal N}}

\def\CD {{\cal D}}
\def\CF {{\cal F}}

\def\CP {{\cal P }}

\def\CO {{\cal O}}

\def\CE {{\cal E}}

\def\CS {{\cal S}}

\def\half{\frac{1}{2}}

\def\one{{\hbox{ 1\kern-.8mm l}}}

\def\vol{{\rm vol\,}}
\def\p{\partial}

\def\be{\bar{e}}

\def\half{\frac{1}{2}}

\def\fe{\mathfrak{e}}

\def\fp{\mathfrak{p}}

\def\fp{\mathfrak{p}}

\def\fE{\mathfrak{E}}

\def\fP{\mathfrak{P}}

\def\half{\frac{1}{2}}

\def\be{ \begin{equation} }
\def\ee{ \end{equation}}

\def\I{{\rm i}}

\title{Computation Of Some Zamolodchikov Volumes, With An Application}

\author{ Gregory W. Moore, \\
 NHETC and Department of Physics and Astronomy,
Rutgers University,\\
Piscataway, NJ 08855--0849, USA\\
$^3$ School of Natural Sciences, Institute for Advanced Study, \\
Princeton, NJ 08540, USA\\
\\
{\rm  gmoore@physics.rutgers.edu } }

\abstract{We compute the Zamolodchikov volumes of some moduli spaces of conformal
field theories with target spaces K3, T4, and their symmetric products. As an application we argue
that sequences of conformal field theories, built from products of such symmetric products, almost never
have a holographic dual with weakly coupled gravity.      August 6, 2015 }

\begin{document}

\section{Introduction And Motivation}

Physicists often speak of the set of all two-dimensional
conformal field theories as a ``moduli space of conformal field theories.''
It is believed that one can make
sense of this set  as a topological
space and that, moreover, the generic point in this space
has a smooth neighborhood locally modeled on a manifold.   It was noted by Zamolodchikov that
at such smooth points the moduli space has a \emph{canonical metric}  \cite{Zamolodchikov:1986gt}. In many
cases of interest in string theory these moduli spaces have
finite volume in the Zamolodchikov metric, a fact of some
importance when one applies statistical ideas to string
compactification \cite{Ashok:2003gk,Douglas:2005hq,Douglas:2006zj,Douglas:2006es,Horne:1994mi}.
In a recent paper \cite{Benjamin:2015hsa} the finiteness of the Zamolodchikov
volume for certain families of 2d CFT's has again played
an important role. The present note should be regarded as an addendum to
\cite{Benjamin:2015hsa}.

In this note we compute the Zamolodchikov volumes for
the 2d nonlinear sigma model whose target space is
(the hyperk\"ahler resolution of) a symmetric product
of K3 surfaces or four-dimensional tori. Moreover, we return to the statistical
considerations of \cite{Benjamin:2015hsa} and show in a
rather precise sense that the set of sequences of
conformal field theories with weakly coupled gravitational
holographic duals, when drawn from the ensembles described in
Section \S \ref{sec:Ensembles} below, is of measure zero.

\section{Zamolodchikov Metric}

If  a two-dimensional conformal field theory $\CC$ is a smooth point in a
moduli space $\CM$ of CFT's then there is a canonical isomorphism
between the tangent space $T_{\CC}\CM$ and the vector space $V^{1,1}$ of
exactly marginal $(1,1)$ operators in $\CC$.
For small $\epsilon$ the
isomorphism takes $\epsilon \CO\in V^{1,1}$ to the tangent vector defined by deforming correlation
functions to:
\be
\left\langle \prod_i \Phi_i \right\rangle \to
\left\langle e^{- \int \epsilon \CO } \prod_i \Phi_i \right\rangle
\ee

In the case where the CFT's are defined by an action we can be a bit more
precise. We define the isomorphism
\be
\Psi: V^{1,1} \rightarrow T_{\CC}\CM
\ee
as follows: If $S[t]$ is a one-parameter family of actions of conformal field theories
(hence a path in $\CM$)
and $\frac{d}{dt}\vert_0 S[t] = \int \CO $ then
\be
\Psi(\CO) =  \frac{\p}{\p t}\biggr\vert_0 .
\ee
In these terms, the   Zamolodchikov metric is then defined by saying that if $v = \Psi(\CO)$ then
\be
\langle \CO(z_1) \CO(z_2) \rangle := g_{\rm Z}(v,v) \frac{d^2z_1 d^2 z_2}{  \vert z_1 - z_2 \vert^4 }
\ee
where the LHS is the correlation function on the complex plane $\IC$ with the unique $SL(2,\IC)$
invariant vacuum at $z=0,\infty$, and   $d^2 z:= dx dy = \frac{\I}{2} dz \wedge d\bar z$
with $z=x+ \I y$. We will also denote the Zamolodchikov metric by $ds^2_Z$.
Note that the Zamolodchikov metric has a \underline{canonical} normalization. When the moduli space
has a finite volume it is therefore meaningful to ask what that volume is.
\footnote{Of course, in some physics problems, a
different normalization of the metric might be called for.
The situation is similar to that of defining an Ad-invariant
metric on a simple Lie algebra. The Ad-invariant metrics are
unique up to scale. There is, however, a canonical normalization - given
by the trace in the adjoint representation. In a metric $\lambda^2 ds^2_Z$ the
volumes quoted in this paper are rescaled by a factor of $\lambda^{\CD}$, where
$\CD$ is the real dimension.}

As an example, consider the
Gaussian model for a periodic real scalar. The action for this model is
\be
S = K  \int d\phi*d\phi
\ee
with $\phi$ dimensionless and periodic (the period can be anything, so long as it is fixed)
and $K$ is a positive constant.
One standard normalization in physics is to take  $\phi \sim \phi + 2\pi$ and
$K = \frac{R^2}{4\pi \alpha'}$.
One readily computes that for the Gaussian model
\be
ds^2_{\rm Z} =  \frac{1}{(2\pi)^2} \left( \frac{dK}{K} \right)^2 = \frac{1}{\pi^2} \left( \frac{dR}{R} \right)^2
\ee
 The factor $ \frac{1}{(2\pi)^2} $
comes from the basic fact that the Green's function in two dimensions is $\frac{1}{4\pi} \log\vert z_1 - z_2 \vert^2$
so $\langle \phi(z_1) \phi(z_2) \rangle = \frac{1}{2\pi K} \log\vert z_1 - z_2 \vert^2$. The moduli space is a half-line,
and has an infinite volume.

Despite this unpromising beginning, the moduli spaces of chiral bosons  in general do have finite volume
moduli spaces.
\footnote{Actually, we should use the term ``moduli stacks,'' but neither the
reader nor the author will need to understand this term to make sense of this paper. }
We consider a theory of   $r$ right-moving   and $r+8s$ left-moving chiral bosons based on even unimodular
lattices $L_{r,s}$ of signature $(+1^{r+8s}, -1^r)$ with $r>0$ and $s \geq 0$. For concreteness,
choose the quadratic form:
\be
Q_{r,s} := U^{\oplus r} \oplus Q_8^{\oplus s}
\ee
where
\be
U = \begin{pmatrix} 0 & 1 \\  1 & 0 \\ \end{pmatrix}
\ee
and $Q_8$ is the Gram matrix of the $E_8$ lattice in some basis of
simple roots. These theories come in moduli spaces
 generally called ``Narain moduli spaces'' in the string theory
literature.
Mathematically, they are the moduli spaces of embeddings of
the lattices $L_{r,s}$
into the pseudo-Euclidean space $\IR^{r+8s,r}$. These moduli spaces can be
expressed as the double-coset
\be
\CN_{r+8s,r} := O_{\IZ}(Q_{r,s}) \backslash O_{\IR}(Q_{r,s})/O(r+8s)\times O(r)
\ee
with
\be
\begin{split}
O_{\IZ}(Q_{r,s}) & := \{ g\in GL(D,\IZ) \vert g^{tr} Q_{r,s} g = Q_{r,s} \} \\
O_{\IR}(Q_{r,s}) & := \{ g\in GL(D,\IR) \vert g^{tr} Q_{r,s} g = Q_{r,s} \}  \cong O(r+8s,r)\\
\end{split}
\ee
where $D :=2r + 8s :=2d$. Note that they have
dimension
\be
\CD = \dim \CN_{r+8s,r} = r (r+8s).
\ee

For Narain moduli spaces the Zamolodchikov metric is a
homogeneous metric induced by analytic continuation of
a left-right invariant metric on the Lie algebra $so(D)$.
This follows since the conformal field theories form an
equivariant bundle of vertex operator algebras over $\CN_{r+8s,r}$
 (See, for examples, \cite{Moore:1993zc,Ranganathan:1993vj}
for the case $s=0$.) The precise normalization  of this
homogeneous metric is of importance to us, and we determine it
as follows:

We first recall how the Narain moduli space is parametrized
in terms of the physical data of a conformal field theory with
toroidal target space of dimension $r$: The physical data consists
of a  flat metric, a flat $B$-field, and a flat gauge connection
for an $E_8^s$ gauge group. We now relate these data to the
embedding of $L_{r,s}$ into  $\IR^{r+8s;r}$. (At this point
we no longer distinguish between $L_{r,s}$ and its embedded version.)
The projection of a vector $p\in L_{r,s}$
to the components in the definite subspaces is denoted $p=(p_L;p_R)$.
Therefore, the vectors can be written as:
%
%
\be
\begin{pmatrix} p_L \\ p_R \\ \end{pmatrix} = \CE \vec n
\ee
where $\CE $ is a $ D \times D$ matrix and $\vec n$ is a $D$-component
integral column vector representing the vector $p$ in an integral basis
of $L_{r,s}$.  We   have
\be\label{eq:InnProds}
\CE^{tr} Q_0 \CE = Q
\ee
where
\be
Q_0 = \begin{pmatrix} 1_{r+8s} & 0 \\ 0 & -1_r \\ \end{pmatrix}
\qquad\qquad
Q = \begin{pmatrix}
Q_{8s} & 0 & 0 \\
0 & 0 & 1_r \\
0 & 1_r & 0 \\ \end{pmatrix}
\ee
and $Q_{8s}$ is the Gram matrix, in some basis,  
of some positive definite even unimodular lattice $\Gamma_{8s}$
of dimension $8s$ (it does not matter which basis and which lattice).
It is a positive definite symmetric even integral matrix of determinant one.
Moreover, there exists $S = S^{tr} = S^{-1}$ so that
\be
Q = \CS^{tr} Q_0 \CS
\ee
where
\be
\CS = \begin{pmatrix} f & 0 \\  0 & S \\ \end{pmatrix}
\ee
and $f$ is a generating
matrix for $Q_{8s}$, that is, a matrix formed from basis vectors for
an embedding of $\Gamma_{8s}$   into the Euclidean space
$\IR^{8s}$ with Gram matrix $Q_{8s}$:
\be
f^{tr} f = Q_{8s}.
\ee
Note that $\CE \CS^{-1} \in O_{\IR}(Q_0)$ and $\CS^{-1}\CE \in O_{\IR}(Q)$
so $\CM$ is diffeomorphic to an orthogonal group.

Denote the space of solutions to \eqref{eq:InnProds} by $\CM$.
A nice parametrization of the homogeneous space
$O(r+8s)\times O(r) \backslash \CM$  is given by a set of representatives  $\CE$
derived from the  zeromodes for the left- and right-moving
chiral fields using formulae from   \cite{Ginsparg:1986bx,Narain:1986am}.
The result is the following:
Let $e_1, e_2$ be two invertible $r\times r$ matrices so that
$e_1^{tr} e_1 = e_2^{tr} e_2 $ is a positive definite symmetric
matrix. Call it $ G^{-1}$. Now let  $B$ be an arbitrary $r\times r$
antisymmetric matrix and form $E:=G+B$. Let $a$ be an
arbitrary $8s \times r$ matrix.
All the matrices $e_1,e_2, B, a, f$ are over the real numbers.
Now consider
\be
\CE = \begin{pmatrix}
 f & 0 &  a \\
 -\half e_1 a^{tr} f & \half e_1 & e_1 (E - \frac{1}{4}   a^{tr} a) \\
  -\half e_2 a^{tr} f & \half e_2 & - e_2 (E^{tr} + \frac{1}{4}   a^{tr} a) \\
  \end{pmatrix}
\ee
The reader can readily check that this
solves \eqref{eq:InnProds}. The form of this expression is preserved by
left-multiplication by $O(8s)\times O(r) \times O(r)$ and the map
from $\CE$ to $[\CE]$ is a surjection onto   $(O(r+8s) \times O(r) \CM$.

Now we consider the subspace   defined by
$a=0$, $B=0$ and $e_1 = e_2 = {\rm Diag}\{ R_1^{-1}, \dots, R_r^{-1} \} $
where $R_i>0$ are the radii of the square torus. The pullback of the
Zamolodchikov metric to this subspace should be the Zamolodchikov
metric for a product of $r$ Gaussian models with radii $R_i$.
%
%
%
%
%
%
%
%
%
%
%
%
Then we simply have
\be
\Tr_{D} \CE^{-1} d \CE \otimes  \CE^{-1} d \CE = 2 \sum_{i=1}^d \left( \frac{dR_i}{R_i} \right)^2
\ee
Therefore, on all of $\CN_{r+8s,r }$ the Zamolodchikov metric $ds^2_{\rm Z}$ corresponds to the homogeneous metric induced from
\be\label{eq:rescale}
 \frac{1}{2\pi^{2} } \Tr_{D} \CE^{-1} d \CE \otimes  \CE^{-1} d \CE.
\ee

\section{Some Ensembles Of $(4,4)$-Superconformal Field Theories}\label{sec:Ensembles}

Consider a subspace of the set of conformal field theories,
measurable in the Zamolodchikov volume form.
When the total Zamolodchikov volume of this subspace
is finite we can use the Zamolodchikov metric to define a probability measure on
this set of CFT's.

In this paper we will apply this simple remark to ensembles of
unitary $(4,4)$ superconformal field theories. We begin with a
collection $\fE$ of CFT's written as a disjoint union over
ensembles of definite central charge
\be
\fE = \amalg_{M} \fE_M .
\ee
For each $M$, the ensemble at fixed central charge will be
a disjoint union of connected components:
\be
\fE_M = \amalg_{\alpha} \fE_{M,\alpha}.
\ee
We will be considering ensembles so that the total volume:
\be
\vol(\fE_M) = \sum_{\alpha} \vol(\fE_{M,\alpha})
\ee
is finite for fixed $M$. (The sum over $M$ of the volumes $\vol(\fE_M)$ will not be finite,
but we will nevertheless refer to $\fE$ as an ``ensemble''.)

In order to apply the techniques of Section \S \ref{sec:AppHolog} we will also want the volumes to be multiplicative:
\be\label{eq:VolTimes}
\vol(\CC_1\times \CC_2) = \vol(\CC_1)\vol(\CC_2).
\ee
In general it is not true that the moduli space of a product of CFT's is the
product of the moduli spaces. That is, in general
 $\CM(\CC_1 \times \CC_2) \not= \CM(\CC_1)\times\CM(\CC_2)$.
Toroidal models give simple examples of this inequality.
Following \cite{Benjamin:2015hsa} we define an ensemble of CFT's such that
\eqref{eq:VolTimes} does hold to be a \emph{multiplicative ensemble}.

We will consider three distinct ensembles defined by the collection of   two-dimensional superconformal field
theories with target spaces of the form:
\be\label{eq:ProdTheory}
\left( S^1 X \right)^{n_1} \times \left( S^2 X \right)^{n_2}\times
\cdots \left( S^r X  \right)^{n_r}.
\ee
In the first ensemble we take $X$ to be a
$K3$ surface, in the second $X$ is a  four-dimensional torus $T4$,  and in the third
$X$ can be either $K3$ or $T4$. The notation  $S^m K3 $  means the
hyperk\"ahler resolution of the symmetric product ${\rm Sym}^m(K3)=(K3)^m/S_m$.
Choosing a complex structure compatible with the hyperk\"ahler structure of $K3$, it is the
Hilbert scheme of points ${\rm Hilb}^m(K3)$ endowed with a hyperk\"ahler metric.
 Similarly we would like to consider the
symmetric product of $T4$, but the resulting space has a nontrivial Betti number $b_1$
and would not define a multiplicative ensemble. Thus, to define $S^m T4$
we begin with the hyperk\"ahler resolution
$\pi: {\rm Hilb}^{m+1}(T4) \rightarrow {\rm Sym}^{m+1}(T4)$ and note that
there is a map $f:{\rm Sym}^{m+1}(T4) \rightarrow T4$ given by taking the sum of
the points. Then $S^m T4$ is defined to be the fiber above zero of $f\circ \pi$.
It is a smooth compact simply connected hyperk\"ahler manifold
\cite{Beauville}. (Warning: $S^1 K3=K3$ but $S^1 T4$   does not equal $T4$. Rather
$S^1 T4$ is the Kummer surface derived from $T4$.)   In all three cases
the ensembles are multiplicative. The reason for this, as noted in \cite{Benjamin:2015hsa},
is that for two Calabi-Yau manifolds $X_1,X_2$ with $h^{1,0}=0$ the number of K\"ahler moduli
$h^{1,1}$ and complex structure moduli $h^{n-1,1}$ is additive for the product $X_1 \times X_2$.
\footnote{One proves this last statement using the K\"unneth formula and
the fact that $h^{1,p}(X_i)$ vanishes unless $p=1$ or $p=\dim_{\IC} X_i - 1$.
This in turn is proved from the Lefshetz hyperplane theorem.
This argument certainly leaves room for a finite quotient in the
relation between $\CM(\CC_1 \times \CC_2)$ and $\CM(\CC_1)\times \CM(\CC_2)$.
The presence of such finite quotients would considerably complicate the
methods of Section \S \ref{sec:AppHolog}, but we believe that it will not
materially affect the main conclusions of that section. In any case, we leave this as an issue
that deserves further thought.}

For $S^1 K3$ the moduli space is just the $80$-dimensional
  space $\CN_{20,4}$.
For $S^M K3$ with $M>1$ and $S^M T4$ with $M \geq 1$
the moduli space can be derived using the attractor mechanism,
as pointed out in \cite{Dijkgraaf:1998gf,Seiberg:1999xz}. We consider
the subgroups of $O_{\IR}(Q_{r,s})$ and  $O_{\IZ}(Q_{r,s})$ fixing a primitive
vector $u$ with $u^2 = 2M$. There is only one such vector
up to equivalence, by the Nikulin embedding theorem \cite{MirandaMorrison,Nikulin}.
Therefore, the conjugacy class of the subgroup only depends
on $M$ and, by abuse of notation, we denote a particular subgroup by
$O_{\IR}(Q_{r,s},M)$, $O_{\IZ}(Q_{r,s},M)$, respectively. Then we have
\be\label{eq:HighLevMod}
\CM( S^M X ) \cong O_{\IZ}(Q_{r,s},M)\backslash O_{\IR}(Q_{r,s},M)/O(r+8s)\times O(r-1)
\ee
where $(r,s)= (5,0)$ for $X=T4$ and $(r,s) = (5,2)$ for $X=K3$.
Note that $O_{\IR}(Q_{r,s})\cong O(r+8s,r-1)$ and so the real dimension is
\be
\dim \left( \CM( S^M X ) \right) = (r-1)(r+8s)
\ee
The space $\CM( S^M X )$
has   real dimension $\CD = 20$ for $X=T4$ and $\CD = 84$ for $X=K3$. The ``extra'' four dimensions
relative to the moduli spaces for $T4$ and $K3$, respectively,
can be thought of as arising from the hypermultiplet of blow-up modes of the locus of $A_1$ singularities
along the big diagonal (where some pair of points coincides) in the symmetric product orbifold.
Thus a neighborhood $U$ in $\CM(X)$ determines a corresponding neighborhood $\cong U \times \left(\IR^3 \times S^1 \right)$
in $\CM(S^M X)$ for $M>1$ where the factor of $\IR^3$ represents levels of three hyperk\"ahler moment
maps and the $S^1$ is a period of a $B$-field.
\footnote{The case of $S^1T4$ deserves one further comment. In general, resolving the $16$ fixed points of $T4/\IZ_2$
results in $16\times 4$ conformal field theory moduli for a total of $16 \times 5 = 80$ moduli. However,
in $\CM(S^1 T4)$ we preserve translation invariance in the resolution, so there are only
$16 +4 =20$ conformal field theory moduli.}

Let $v^{\rm tr}_X(M)$ denote the volume of the moduli space for $S^M X$
in the homogeneous metric induced from $ds^2 = \Tr_D (\CE^{-1} d\CE)^2$.
According to equation \eqref{eq:rescale} the Zamolodchikov metric is related 
by rescaling with the factor $1/(2\pi^2)$ and
so volumes are rescaled by  $1/(\pi\sqrt{2})^{\CD}$  where $\CD$ is
the dimension of the moduli space.   In addition, in a symmetric product
orbifold with $M$ factors the Zamolodchikov metric is also rescaled by a factor of $M$.
We justify this last statement as follows:

In general there is an immersion (most likely, an embedding) of moduli spaces
\be\label{eq:ModImmerse}
\iota: \CM(\CC) \hookrightarrow \CM({\rm Sym}^M(\CC)).
\ee
Indeed, if a point in $\CM(\CC)$ has action $S$ then the
action for the corresponding point in $\CM({\rm Sym}^M(\CC))$ is just
\be
S^{(1)} + \cdots + S^{(M)},
\ee
where $S^{(i)}$ is the action $S$ for the $i^{th}$ factor. 
Note that there is no overall factor involving a power of $M$.
(This last statement can be verified by noting that the stress-energy tensor of ${\rm Sym}^M(\CC)$ is the sum of the stress-energy
tensors of $\CC$. Recall then that the energy-momentum tensor is the variation of the action with respect to
the worldsheet metric.) This immersion is simple enough, when viewed from the CFT
perspective, but is rather more nontrivial when understood in terms of moduli spaces of
hyperk\"ahler metrics.  In any case, we claim that, when restricted to the image of $\iota$ in \eqref{eq:ModImmerse}
we have the commutative diagram:
\be
\xymatrix{ V^{1,1}({\rm Sym}^M(\CC)) \ar[r]^{\Psi} & T \CM({\rm Sym}^M(\CC))\\
V^{1,1}(\CC)\ar[u]^{\varphi}\ar[r]^{\Psi} & T\CM(\CC) \ar[u]^{\iota_*} \\ }
\ee
where
\be
\varphi(\CO) = \CO^{(1)} + \cdots  + \CO^{(M)}.
\ee
Again, the absence of a normalization factor involving a power of $M$
 follows from the definition in terms of the deformation of the action, together with the additivity of the
actions noted above.  Therefore,
\be
\iota^*(ds^{2}_{\rm Z}( {\rm Sym}^M(\CC))) = M ds^2_{\rm Z}(\CC).
\ee

In our application to $\CM(S^M X)$, the metric is again a homogeneous metric, so the scale is
determined by restricting to the subspace isomorphic to $\CM(S^1 X)$. Letting $v_X^{\rm Zam}(M)$
denote the volume in the Zamolodchikov metric we have:
\be\label{eq:Vols}
v_X^{\rm Zam}(M)  =  \begin{cases}
\left( \frac{M}{2\pi^2} \right)^{42} v_X^{\rm tr}(M) & X=K3,\quad M> 1 \\
\left( \frac{M+1}{2\pi^2} \right)^{10} v_X^{\rm tr}(M) & X=T4,\quad M \geq 1 \\
\end{cases}.
\ee

We will return to these three ensembles, and their Zamolodchikov measures in
 Section \ref{sec:AppHolog} below.

\section{Some Results From Number Theory}

\subsection{Volumes For $\CN_{r+8s,r}$}

The volumes for the spaces $\CN_{r+8s,r}$ are related to the so-called
``mass'' of a genus of lattices determined by $Q_{r,s}$.
\footnote{The word ``mass'' is a mistranslation of the German word ``mass.''
The correct translation is ``measure,'' a term that makes much better sense. But the term
``mass'' has become standard.}
 The
``mass'' of a genus of lattices
was introduced in the work of Carl Ludwig Siegel.  In
the case with $r>0$ there is only one equivalence class in the
genus and the mass can be identified as the volume of a fundamental domain:
\be
Mass(L_{r,s})=   \int_{O_{\IZ}(Q_{r,s} )\backslash O_{\IR}(Q_{r,s} )} \mu
\ee
where $\mu$ is a left-right invariant measure. Siegel gave general formulae
for this and related expressions. Siegel's formulae involve products of ``local density factors''
over all prime numbers, and the computation of those densities is itself nontrivial.
In the papers \cite{BG,GHY} the relevant factors were computed for
the general case of unimodular lattices (not necessarily even). The paper
\cite{BG} gives the volume using the measure $ \mu= \mu^{\rm cpt}$ normalized so that
its analytic continuation to the connected  compact group $SO(D)$ gives unit total volume:
\be\label{eq:cpt-norm}
\vol^{\rm cpt}(SO(D))=1
\ee
where $D=2r+8s =2d$.
\footnote{Reference \cite{BG} does not state whether the ``compact form'' is
$O(D)$ or $SO(D)$. The computation for the special case $r=2,s=0$ in the appendix
suggests that $SO(D)$ is what was meant. This interpretation has kindly been confirmed by
M. Belolipetsky.  }
In particular, applying Theorem 3.1 of \cite{BG} to our case gives
\be\label{eq:Mass-Lrs}
Mass(L_{r,s}) = 2 (d-1)! \frac{\zeta(d)}{(2\pi)^d}  \prod_{j=1}^{d-1} \frac{ \vert B_{2j} \vert}{4j} .
\ee
Readers who wish to check this specialization should   note that
the Tamagawa number $\tau(G)=2$ since $G$ is an orthogonal group,
$d_F=1$ and   $\deg(F)=1$ since $F=\IQ$, and likewise   $disc(q)=+1$.
For the local factors   $\lambda_p$
for all the odd primes $p$ we consult Table 3, p. 117 of \cite{GHY}. We are
in the case $\delta =1$ and the Hasse-Minkowski-Witt invariant $w=1$, because
all the Hilbert symbols are $(\pm 1, \pm 1) = 1$. Therefore we have the
first line of the table and $\lambda_p=1$. For the prime $p=2$ we consult
Corollary 5.3 of \cite{BG} and again, $\lambda_{p=2}=1$. Finally, we
used $\zeta(2j)/(2\pi)^{2j} = \vert B_{2j}\vert/(2 (2j)!)$ to rewrite the
formula slightly.

Now we have to convert to the measure $\mu^{\rm tr}$. We do this by
noting that $O(n+1)/O(n)\cong S^n$, the $n$-dimensional sphere. The
sphere $S^n$ of radius $R$  has volume
\be
R^n \frac{2\pi^{(n+1)/2}}{\Gamma((n+1)/2)}
\ee
in the standard round metric. A small computation shows that the homogeneous
metric induced from $-\Tr_{n+1} (g^{-1} dg)^2$ gives the round metric with
radius $\sqrt{2}$ and hence
\be
\vol^{\rm tr}(O(D)) := \sigma(D) = 2^{(D+1)/2} \prod_{j=1}^{D-1} \left( \frac{(2\pi)^{\frac{j+1}{2}}}{\Gamma( \frac{j+1}{2} )}\right)
\ee
where $\vol^{\rm tr}(O(D))$ is the volume in the metric $-\Tr_{D} (g^{-1} dg)^2$.

Combining the above remarks we have:
\be\label{eq:NarainVols}
\vol^{\rm tr}(\CN_{r+8s,r}) = \frac{\sigma(2r+8s)}{\sigma(r)\sigma(r+8s)}  \cdot
2 (d-1)! \frac{\zeta(d)}{(2\pi)^d}  \prod_{j=1}^{d-1} \frac{ \vert B_{2j} \vert}{4j}
\ee
In arriving at this formula there are two canceling factors of two. We have multiplied by
the volume of $O(D)$ (giving an ``extra'' factor of two). However the element
$(-1,-1)\in O(r) \times O(r+8s)$ does not act effectively on the quotient $O_{\IZ}(Q_{r,s} )\backslash O_{\IR}(Q_{r,s} )$,
so when dividing by  $\sigma(r)\sigma(r+8s)$ we have divided by an ``extra'' factor of two.
\footnote{In \cite{Ashok:2003gk} Ashok and Douglas cited a related computation of
Siegel for the volume of the moduli space of complex structures on a torus.
The above formula for the volume of $\CN_{r,r}$ differs from the formula they use since the K\"ahler modes and $B$-fields
are not included in their formula.   }

%
%
%
%
%

\subsection{Volumes For   $S^M X$ }

Now we turn to the rather more challenging case of computing
\be
\vol\left( O_{\IZ}(Q_{r,s},M)\backslash O_{\IR}(Q_{r,s}, M) \right)
\ee
where the notation
was defined above equation \eqref{eq:HighLevMod}.
For applications to the ensembles of Section \S \ref{sec:Ensembles}
we will specialize to
$r=5$ and $s=2$ and $r=5$ and $s=0$.

The paper of C.L. Siegel \cite{Siegel-IndefiniteForms} computes the
volume of $O_{\IZ}(Q_{r,s},M)\backslash O_{\IR}(Q_{r,s}, M)$
 in a measure $\mu^{\rm cls}$ normalized so
that
\be
\vol^{\rm cls}(O(D)) = \prod_{j=1}^{D} \frac{\pi^{j/2}}{\Gamma(j/2)}.
\ee
Therefore we will have to take into account a fudge factor
\be\label{eq:MoreFudge}
\vol^{\rm tr}(O(D)) = 2^{D(D+3)/4} \vol^{\rm cls}(O(D)).
\ee
As long as $r>1$, so that there is only one class of primitive vector $u$,
the Siegel formula reduces to:
\be\label{eq:SiegelFormula}
\frac{ \vol^{\rm cls}( O_{\IZ}(Q_{r,s},M)\backslash O_{\IR}(Q_{r,s}, M)) }{\vol^{\rm cls}( O_{\IZ}(Q_{r,s}) \backslash O_{\IR}(Q_{r,s} ))} = \prod_{p<\infty} \alpha_p(M)
\ee
where the product is over all the finite primes,
\be
\alpha_p(M):= \lim_{t\to \infty} \frac{A(d,M,p^t) }{p^{t(2d-1)}},
\ee
and $A(d,M,p^t)$ is the number of representatives of $2M$ by $Q_{r,s}$ over the ring $\IZ/p^t\IZ$. That is:
\be\label{eq:Adp-def}
A(d,M,p^t) = \# \{ v~ \mod p^t \vert Q_{r,s}(v) = 2M ~ \mod p^t \}
\ee
We now evaluate the ``local densitites'' $\alpha_p(M)$. To do this, we need a formula for $A(d,M,p^t)$.

The first remark is that the lattice $L_{r,s}$ is equivalent over the $p$-adic integers, for all finite
primes $p$,  to the simpler lattice $U^{\oplus d}$. For an odd prime $p$ this
follows, for example, from Theorem 3.1, p.115 of Cassels' book \cite{Cassels}.
(Alternatively, one can apply Theorems 2 and 9 of Chapter 15 of \cite{SPLG}.)
The prime $p=2$ is more delicate. Using the classification over $2$-adic integers
described in \cite{MirandaMorrison,Nikulin} and unimodularity
 we conclude that the forms are also equivalent over
the $2$-adic integers. Therefore, $A(d,M,p^t)$ only depends on  $r,s$ through the combination $d=r+4s$,
as indicated in the notation. Moreover, we can replace \eqref{eq:Adp-def}
with the much simpler expression:
\be\label{eq:Adp-def2}
A(d,M,p^t) = \# \{x_i,y_i \mod p^t \vert  \sum_{i=1}^d 2 x_i y_i  = 2M ~ \mod p^t \}.
\ee
We find this surprising: The only thing standing between
the bland and boring $U^{\oplus d}$ and the arresting and attractive $L_{r,s}$ with
its beautiful $E_8$ summands is the prime at infinity!

Next, we remark that,  for each prime $p$
the answer only depends on the power of $p$ that divides $M$. So it suffices
to compute $A(d,p^e,p^t)$, where $e\geq 0$, and we only consider $t>e$.
We now have the result that for $p$ an \emph{odd} prime:
\be\label{eq:NiceAnswer}
A(d,p^e,p^t) = p^{(2d-1)t } (1-p^{-d} ) \frac{1- p^{-(e+1)(d-1)} }{1- p^{-(d-1)} }
\ee
while for $p=2$ we have
\be\label{eq:NiceAnswer2}
A(d,2^e,2^t) =  2\cdot 2^{(2d-1)t } (1-2^{-d} ) \frac{1- 2^{-(e+1)(d-1)} }{1- 2^{-(d-1)} }
\ee
These formulae also apply for $d=1$ if we interpret the final quotient using L'Hopital's rule.

We now prove \eqref{eq:NiceAnswer} and \eqref{eq:NiceAnswer2}.

\subsubsection{Proof Of The Formula For $A_p(d, p^e, t)$ }

To begin, we consider the case of an odd prime. Let us write  \eqref{eq:NiceAnswer} as
\be\label{eq:NiceAnswer3}
p^{(2d-1)(t-e-1) + de + (d-1) }  (p^d-1) \bigl[ 1 + p^{d-1} + (p^{d-1} )^2 + \cdots + (p^{d-1})^e \bigr]
\ee
This makes it clear that it is an integer in the range $t>e$ where it is meant to hold.
We divide the set of solutions $\CS$ into two disjoint sets $\CS = \CS_1 \amalg \CS_2$
where $\CS_1$ is the set of solutions where  at least one of the $x_i$ is invertible mod $p^t$,
while $\CS_2$ is the set of solutions where all of the $x_i$ fail to be invertible mod $p^t$.

We first show that the number of solutions  in $\CS_1$ is:
\be
 p^{t(2d-1)}(1-p^{-d}).
\ee
To do this we further divide up $\CS_1$ into a disjoint union of sets $\CS_1^j$,
with $j=1,\dots, d$. The set $\CS_1^j$ is the set of solutions such that $x_j$ is
  invertible mod $p^t$, but $x_i$ for $i<j$ are not invertible mod $p^t$.
The number of invertible elements in $\IZ/p^t\IZ$ is $p^t-p^{t-1}$ and the number of
noninvertible ones is $p^{t-1}$. When $x_j$ is invertible, we can solve for $y_j$.
Therefore, the total number of solutions in $\CS_1$ is just
\be
(p^t - p^{t-1}) (p^t)^{d-1} \sum_{j=1}^{d} (p^{t-1})^{(j-1)} (p^t)^{d-j} = p^{t(2d-1)}(1-p^{-d})
\ee
where $(p^t - p^{t-1})$ is the number of invertible choices for $x_j$, $(p^{t-1})^{(j-1)}$ is the number
noninvertible choices for $x_i$ with $i<j$, $(p^t)^{d-j}$ is the number of arbitrary
choices for $x_i$ with $i>j$,  and $(p^t)^{d-1}$ is the number of choices for $y_k$, with $k\not=j$.
Note that for the case $e=0$ the set of solutions $\CS_2$ is empty, since then $2M$ is prime to $p$, so we have
 proven \eqref{eq:NiceAnswer3} for $e=0$.
\footnote{The argument can be also be used to
count the number of solutions to $\sum_i x_iy_i=0 \mod p$ because the only
other case is where all $x_i=0$. But then $y_i$ can be anything, so we just add $p^d$,
thus getting $p^{2d-1} - p^{d-1} + p^d$ solutions.}

We can now prove the general result by induction on $e$. If $e>0$ the set $\CS_2$ will be nonempty.
If   all the $x_i$ fail to be invertible then we can write $x_i = p \tilde x_i$ where $\tilde x_i$ is
defined mod  $p^{t-1}$. Moreover,
\be
\sum_i \tilde x_i y_i = p^{e-1} \mod p^{t-1}
\ee
By the inductive hypothesis we know the number of solutions, modulo $p^{t-1}$, to this equation is
\be
p^{(2d-1)(t-e-1) + d(e-1) + (d-1) }  (p^d-1) \bigl[ 1 + p^{d-1} + (p^{d-1})^2 + \cdots + (p^{d-1})^{e-1} \bigr]
\ee
But now we need to lift the solutions $(\tilde x_i, y_i)\mod p^{t-1}$ to solutions mod $p^t$.
The lifts of $\tilde x_i$ are $\tilde x_i + a_i p^{t-1}$ and  do not change the value of $x_i = p \tilde x_i \mod p^t$.
Moreover, \emph{all} the lifts of $y_i \to y_i + b_i p^{t-1}$ solve the equation because
$x_i (y_i + b_i p^{t-1}) = x_i y_i + \tilde x_i b_i p^t = x_i y_i \mod p^t$. So all $p^d$ lifts of the vector
$(y_1,\dots, y_d)$ produce solutions. So the number of solutions of type $\CS_2$ is
\be
p^{(2d-1)(t-e-1) + de + (d-1) }  (p^d-1) \bigl[ 1 + p^{d-1} + (p^{d-1})^2 + \cdots + (p^{d-1})^{e-1} \bigr]
\ee
Combining this with the number of solutions of type $1$ we arrive at the desired \eqref{eq:NiceAnswer3}.

Now, turning to the case of $p=2$, we note that if
\be\label{eq:RedArg1}
\sum_{i=1}^d 2 x_i y_i = 2\cdot 2^e \mod 2^t
\ee
(where we recall that $M=2^e u$, with $u$ odd and $e\geq 0$)
then
\be\label{eq:RedArg2}
\sum_{i=1}^d   x_i y_i = 2^{e } \mod 2^{t-1}
\ee
Moreover, given a solution of \eqref{eq:RedArg2} with $x_i, y_i$ defined modulo $2^{t-1}$ there are $2^{2d}$
distinct lifts of solutions to \eqref{eq:RedArg1} modulo $2^t$. On the other hand, the argument we gave
for the odd primes works equally well for counting solutions of \eqref{eq:RedArg2}.
(The only slightly subtle point is that the group of invertible elements of $\IZ/2^t \IZ$ is not cyclic for $t>1$.
Nevertheless, it is still of order  $2^{t}-2^{t-1} = 2^{t-1}$,
and that is all we used.)

\subsubsection{Answer For The Volumes}

Given the formulae \eqref{eq:NiceAnswer} and \eqref{eq:NiceAnswer2} we get
the local densities:
\be
\alpha_p(M) = \begin{cases}  (1-p^{-d}) \frac{ 1- p^{-(e_p(M)+1)(d-1)}}{1-p^{-(d-1)}} & p \not=2  \\
 2\cdot (1-p^{-d}) \frac{ 1- p^{-(e_p(M)+1)(d-1)}}{1-p^{-(d-1)}} & p =2 \\
\end{cases}
\ee
where $e_p(M)$ is the $p$-adic valuation of $M$. That is: $M=\prod_{p} p^{e_p(M)}$.

The formula \eqref{eq:SiegelFormula} becomes
\be\label{eq:Rat}
\frac{ \vol^{\rm cls}( O_{\IZ}(Q_{r,s},M)\backslash O_{\IR}(Q_{r,s}, M)) }{\vol^{\rm cls}(O_{\IZ}(Q_{r,s}) \backslash O_{\IR}(Q_{r,s}))} =
2 \zeta(d)^{-1} f_d(M)
\ee
where
\be\label{eq:fd-def}
f_d(M) =      \prod_{p\vert M} \frac{ 1- p^{-(e_p(M) +1)(d-1)}}{1-p^{-(d-1)}}
\ee

Taking into account the fudge-factor going from $\vol^{\rm cls}$ to $\vol^{\rm tr}$ and
defining
\be
V_{r,s}(M) := \vol^{\rm tr}( O_{\IZ}(Q_{r,s},M)\backslash O_{\IR}(Q_{r,s}, M)/( O(r+8s)\times O(r-1) ) )
\ee
we conclude that:
\be\label{eq:VolrsM}
V_{r,s}(M) = \sqrt{8}   \frac{\sigma(2r+8s)}{\sigma(r-1)\sigma(r+8s)} \frac{(d-1)!}{(4\pi)^d}
\left( \prod_{j=1}^{d-1} \frac{\vert B_{2j}\vert}{4j}\right)  f_d(M).
\ee

\section{Application To Holography  }\label{sec:AppHolog}

\subsection{General Strategy}

In this subsection we review some of the considerations from
Section VI of \cite{Benjamin:2015hsa}.  For more background  see \cite{Benjamin:2015hsa}.
The considerations of \cite{Benjamin:2015hsa} were motivated by the following question:

\emph{ Consider a sequence $\CC_M$ of conformal field theories (say, with $(4,4)$ supersymmetry
and $c=6M$).  ``How likely'' is it that this sequence has a large $M$ holographic dual with weakly coupled
gravity? }

Following the recent papers \cite{Belin:2014fna,Hartman,Keller:2011xi},
reference \cite{Benjamin:2015hsa} proposed that an important
\emph{necessary} criterion for the existence of such a holographic dual is that the
elliptic genus should exhibit a Hawking-Page phase transition.  Reference \cite{Benjamin:2015hsa}
further argued that a necessary criterion for a Hawking-Page phase transition is that
the absolute value of the extremal polar coefficient, denoted
$\fe(\CC_M)$, must grow at most polynomially in $M$, that is,  it is $o(\exp(M^\delta))$ for
any $\delta>0$.

In principle, we should define a probability measure on \emph{sequences} of conformal field
theories $\{ \CC_M \}$ drawn from the ensembles $\fE$ described in Section \S \ref{sec:Ensembles}.
We will not do that here.
As a surrogate, we will instead state some (possibly $M$-dependent) property
$\CP$ of a CFT and instead consider sequences of probabilities $p_M$ that CFT's of central charge $M$
(drawn from the ensemble $\fE_M$) satisfy property $\CP$. For example, $\CP$ might be the statement that
$\fe(\CC) \leq \kappa M^\ell$. The measure in $\fE_M$ for $\CC$ to have $\fe(\CC)=\fe$ is
\be
\mu(\fe;M):= \frac{\vol(\fe;M)}{\vol(M)}
\ee
where $\vol(M)=\vol(\fE_M)$ and $\vol(\fe;M)$ is the volume of the subset in $\fE_M$ of
theories with $\fe(\CC)=\fe$. Therefore, we introduce:
\be
p_M(\kappa,\ell) = \sum_{\fe \leq \kappa M^\ell} \frac{\vol(\fe;M)}{\vol(M)}.
\ee
If the limit
\be
\lim_{M\to \infty} p_M(\kappa ,\ell)
\ee
exists and is independent of $\kappa$, we will say that it is the probability that a sequence of
CFT's $\{\CC_M\}$ drawn from the ensemble $\fE$ has $\fe$ growing at most like a power $M^\ell$.
We will denote it by $\fp(\ell)$.

If our ensemble is a multiplicative ensemble then it is useful to  define \emph{prime} CFT's to be
those which are not a product (even up to deformation) of CFT's in $\fE$ with positive central
charge.  Denoting the prime CFT's at a fixed
central charge $c=6m$ by $\CC_{m,\alpha}$ we can form the generating functional
\begin{equation}\label{eq:GenFun}
\prod_{m=1}^{\infty}\prod_{\alpha}  \frac{1}{1- v_{\alpha}(m) \fe_{\alpha}(m)^{-s} q^m }
=  1 + \sum_{M=1}^\infty \xi(s;M) q^M
\end{equation}
where $\fe_{\alpha}(m)$ and $v_{\alpha}(m)$ are the extremal polar coefficient and volumes
associated to $\CC_{m,\alpha}$, respectively. The coefficient of $q^M$ is
\be
\xi(s;M)= \sum_{\fe=1}^\infty \frac{\vol(\fe;M)}{\fe^s }
\ee
and it gives the  volumes $\vol(\fe;M)$.
(For later use, note that $\xi(0;M)={\rm vol}(M)$ is the total volume of the theories with
fixed central charge $c=6M$.)

\subsection{Ingredients For The Three Ensembles}

In the three ensembles we are considering the prime CFT's of index $m$
are $S^m K3$, $S^m T4$, and $\{ S^m K3, S^m T4 \}$, respectively.

The extremal polar coefficient for $S^m K3$ is easily deduced
from the formula for symmetric product orbifolds \cite{Dijkgraaf:1996xw}
and is
\be
\fe(S^m K3 ) = m+1 .
\ee
For $S^m T4$ the relevant result can be deduced
from \cite{Maldacena:1999bp} and is again
\be
\fe(S^m T4) = m+1 .
\ee
(We provide a few details in Appendix \ref{App:T4} below.)

Let us denote the Zamolodchikov volumes of $\CM(S^M X)$ by $v_X(M)$.
Recalling equation \eqref{eq:Vols} we have: 
\be
v_{K3}(M) = \begin{cases}
\left( \frac{1}{2\pi^2} \right)^{40} \vol^{\rm tr}(\CN_{20,4}) & M=1 \\
\left( \frac{M}{2\pi^2} \right)^{42} V_{5,2}(M) & M>1 \\
\end{cases}
\ee
\be
v_{T4}(M) = \left( \frac{M+1}{2\pi^2} \right)^{10} V_{5,0}(M) \qquad  M\geq 1
\ee
%
%
%
and now, thanks to equations \eqref{eq:NarainVols} and \eqref{eq:VolrsM},
we have explicit results for these volumes.
 The numerical values are amusing. We have
\be
\begin{split}
v_{K3}(1)  & = \frac{(131)(283)(593)(617)(691)^2(3617)(43867)}{2^{40}\cdot
3^{34} \cdot 5^{15} \cdot 7^{9}\cdot 11^{5}\cdot 13^{4}\cdot 17^{3}\cdot 19^{3}\cdot 23 \cdot \pi^{40} }
\cong     1.66 \times 10^{-61} \\
\end{split}
\ee
and for $M>1$:
\be
v_{K3}(M) =  \rho M^{42}  f_{13}(M)
\ee
with
\be
f_{13}(M) =      \prod_{p\vert M} \frac{ 1- p^{-12- 12 e_p(M) }}{1-p^{-12}}
\ee
where
\be
\begin{split}
\rho & =
\frac{(103)( 131)(283)( 593)( 617)(691)(3617)(43867)( 2294797)}{
2^{51}\cdot 3^{35}\cdot 5^{15}\cdot  7^{ 10}\cdot  11^{ 5}\cdot  13^{ 4}\cdot 17^{ 3}\cdot 19^{3} \cdot  23^{2}\cdot \pi^{42} }\\
&
\cong   5.815 \times 10^{-63}\\
\end{split}
\ee
Similarly, we have
 %
%
%
\be
v_{T4}(M) =  \rho' (M+1)^{10}  f_{5}(M)
\ee
with
\be
f_{5}(M) =      \prod_{p\vert M} \frac{ 1- p^{-4- 4 e_p(M) }}{1-p^{-4}}
\ee
where
\be
\begin{split}
\rho' & =
\frac{1}{ 2^{14} \cdot 3^{8} \cdot 5^{4}\cdot 7^{2}\cdot \pi^{10}  }\\
&
\cong   3.24\times 10^{-18} \\
\end{split}
\ee

\subsection{Evaluating The Probabiity $\fp(\ell)$}

Let us now return to the evaluation of $\fp(\ell)$ for the three ensembles.
We consider the functions:
\be\label{eq:Hk-def}
H_\ell(s) :=\lim_{M\to \infty}  (M+1)^{\ell s} \frac{\xi(s;M)}{\xi(0;M)} = \lim_{M\to \infty}  \sum_{\fe=M+1}^{2^M}  \frac{\vol(\fe;M)}{\vol(M)}
\left( \frac{(M+1)^\ell}{\fe} \right)^s
\ee
We claim that the limit exists and moverover, for all positive integers $\ell$, it converges to
the characteristic function:
\be\label{eq:chi-def}
\chi(s) = \begin{cases} 1 & s=0 \\ 0 & s> 0 \\   \end{cases}.
\ee
By splitting the sum in \eqref{eq:Hk-def} into terms with $\fe \leq \kappa (M+1)^\ell$ and $\fe> \kappa (M+1)^\ell$
it is easy to see that $(M+1)^{\ell s} \frac{\xi(s;M)}{\xi(0;M)} \geq \kappa^{-s} p_{M}(\kappa,\ell)\geq 0 $,
and hence if $H_\ell(s) = \chi(s)$
it must be that $\lim_{M\to \infty}  p_M(\kappa,\ell) = 0 $  for all $\kappa$ and $\ell$.

It is precisely in this sense that we mean that almost none of the sequences of CFT's drawn from the
the three ensembles defined in Section \S \ref{sec:Ensembles} have weakly coupled holographic duals.

%

\begin{figure}
\begin{center}
\includegraphics[width=0.75\textwidth]{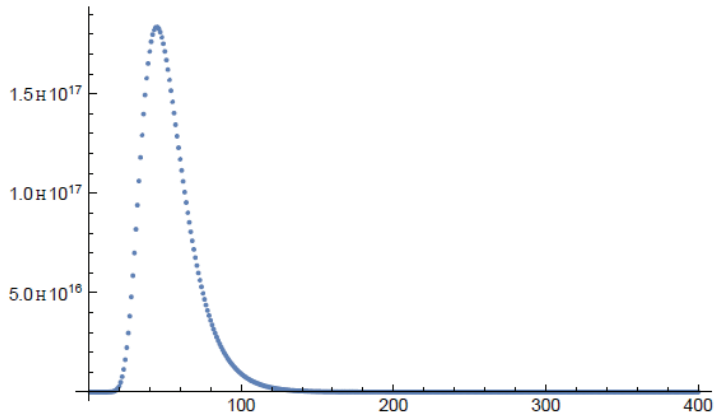}
\end{center}
\caption{
Showing the distribution of $p_k(n)$ as a function of $k$ for $n=400$. The Erd\"os-Lehner mean value of $k$,
$\bar k = \frac{\sqrt{6}}{2\pi} 20 \log(20) \cong 46.7153$,  is a very good approximation
to the location of the maximum of the distribution. The actual maximum is at $k=45$. }
\label{fig:DistributeIntegerParts5}
\end{figure}

%

Our strategy for proving that $H_\ell(s) = \chi(s)$ is to note that we can organize the
terms contributing to $\xi(s;M)$ in terms of partitions of $M$:
\be\label{eq:PartOf-M}
 M = \lambda_1 + \cdots + \lambda_k  \qquad \qquad  \lambda_1 \geq \lambda_2 \geq \cdots \geq \lambda_k .
\ee
Expanding the $N^{th}$ term in the product in \eqref{eq:GenFun} produces the parts in the partition with $\lambda_j = N$.
We now ask: ``What is the ``typical'' partition for large $M$?'' This is an imprecise, and rather
subtle question. To get some sense of an answer it is useful to consider the number $p_k(n)$
of partitions of $n$ into precisely $k$ parts (as in \eqref{eq:PartOf-M}).
A generating function is
\footnote{Note that the transpose of a partition with $k$ parts is a partition whose largest part
has size $k$. The generating function is more obvious from the latter viewpoint.}
\be
\sum_{n=1}^\infty p_k(n)x^n = x^k \prod_{j=1}^k \frac{1}{1-x^j}
\ee
and some naive estimates with Stirling's formula suggests that the distribution of $p_k(n)$
as a function of $k$ for large $n$ should be peaked around $k\cong \sqrt{n}$.
This is, in part, confirmed by looking at numerical data. See, for example, Figure
\ref{fig:DistributeIntegerParts5}.
One natural guess, then, is that the ``typical'' partition has $k\cong \sqrt{n}$
with ``most of the parts'' on the order of $\sqrt{n}$.

The above naive picture can be considerably improved using the statistical theory of
partitions \cite{ErdosLehner,SzalayTuran,VershikYakubovich}.
Without going into a lot of complicated asymptotic formulae, the main upshot is that
$p_k(n)$ indeed is sharply peaked with a maximum around
\be
\bar k(n):= \frac{\sqrt{6}}{2\pi} \sqrt{n}\log n.
\ee
See Figure \ref{fig:DistributeIntegerParts5} for a numerical illustration.
Moreover, according to \cite{SzalayTuran,VershikYakubovich}, and again speaking very roughly,  the number of terms in the
partition $\lambda_j$ with $\lambda_j \cong \frac{\sqrt{6}}{2\pi} \sqrt{n}$
is order $\sqrt{6n}/\pi$.

Very roughly speaking, then,   the  dominant source of partitions of $M$ for large $M$ should
have approximately $\bar k(M)$ parts with most of the parts of order $\sqrt{M}$.
The key fact about $v_X(N)$ we need to know is that the dominant effect
is the power $N^{42}$ for $X=K3$ and $N^{10}$ for $X=T4$. The arithmetic
functions $f_{13}(N)$ and $f_5(N)$ do not change the value significantly.
They are clearly bounded by $N$.
\footnote{A better bound is $f_{d}(N) \leq N^{C_d}$ with
$C_d = \log[2^{d-1}/(2^{d-1}-1)]/\log 2$ giving $C_{13}=0.000352263$
and $C_5 = 0.0931094$. To prove it  note that $f_d(M) \leq \prod_{p\vert M} (1- p^{-(d-1)})^{-1}$,
but $(1- p^{-(d-1)})^{-1} \leq p^{C_d}$ since $\frac{1}{x} \log[e^{(d-1)x}/(e^{(d-1)x}-1)]$
is a monotone decreasing function of $x$.}
Therefore, $v_{K3}(N)$ has a growth larger than $N^{42}$ and smaller than $N^{43}$
and similarly for $X=T4$.  The growth of these volumes
with a large power of $N$ might seem to pose a problem but the
number of partitions with parts on the order of $M$ is relatively exponentially small.
We naively estimate $\xi(s;M)$ by taking   $\bar k(M)$ parts of order $\lambda_j \sim c \sqrt{M}$
and thereby   expect the asymptotics of $\xi(s;M)$ to be given by
\be
\left( v_X(c\sqrt{M})\right)^{\bar k(M)} (c\sqrt{M})^{-s \bar k(M) }e^{2\pi \sqrt{\frac{M}{6}}  } + \cdots
\ee
for some constant $c$. Therefore we expect the asymptotics of $\xi(s;M)/\xi(0;M)$ to be
of the form
\be
\frac{\xi(s;M)}{\xi(0;M)} \sim (c\sqrt{M})^{-s \bar k(M) }
\ee
Therefore, the limit
defining $H_{\ell}(s)$ is given  by
\be
 \lim_{M\to \infty}   (M+1)^{\ell s} (c\sqrt{M})^{-s \bar k(M) } = \chi(s),
\ee
thus establishing our claim for the first two ensembles with $X=K3$ or $X=T4$.

For the third ensemble with $X$ drawn from $\{ K3, T4 \}$ we need to look at
pairs of partitions summing to $M$. When we enumerate such partitions the
sum
\be
\sum_{j=0}^M p(j) p(M-j),
\ee
where $p(n)$ is the orderinary partition function,
has a saddle-point at $j = M/2$ so we expect the previous arguments to give, once
again, $H_{\ell}(s)=\chi(s)$.

The above arguments are admittedly extremely rough. It would be worthwhile to prove
the above claim more carefully.

\section*{Acknowledgements}

GM would like to thank Shamit Kachru for initiating the collaboration
that formed the key background for the above work  \cite{Benjamin:2015hsa}.
He also thanks Miranda Cheng and Shamit Kachru for insisting that he speak at the
Perimeter Workshop on Mock Modular Forms and Moonshine, April 2015. If they had not
done so this paper would never have been written. He also thanks Miranda Cheng for many comments on
the draft and for suggesting that the results should be extended from $K3$ to $T4$.
He finally thanks Christoph Keller, Alex Maloney, Steve Miller, David Morrison, Hiraku Nakajima, Peter Sarnak,
and Edward Witten for useful discussions, and M. Belolipetsky for useful correspondence.
  GM is supported by the DOE under grant
DOE-SC0010008 to Rutgers.   GM also gratefully acknowledges
the hospitality of the Perimeter Institute for Theoretical Physics,
the Aspen Center for Physics  (under
NSF Grant No. PHY-1066293), and finally the Isaac Newton Institute, where this work was completed.

\appendix

\section{An Explicit Computation Of The Volume Of $\CN_{2,2}$ }

One can compute the volume of $\CN_{2,2}$ explicitly
using the fact that $SL(2,\IR) \times SL(2,\IR)$
double-covers the identity component of $O_{\IR}(II^{2,2})$.
(One must keep track of several tricky factors of two in this 
computation.)
 Using the generators of the duality group
$O_{\IZ}(II^{2,2})$ given in \cite{Giveon:1990era} one
can check that the Narain space $\CN_{2,2}$ is a quotient
of a product of upper half-planes $\CF \times \CF$ by
$\IZ_2 \times \IZ_2$, acting via $(\tau, \rho) \rightarrow
(\rho, \tau)$ and $(\tau, \rho ) \rightarrow (- \bar \tau, - \bar \rho)$.
On the other hand, the pullback of the metric $\Tr_{D=4}(\CE^{-1} d \CE)^2 $
is $2( \Tr(A^{-1} dA)^2 + \Tr(B^{-1} dB)^2)$ where $\CE$ pulls back to
a pair $(A,B) \in SL(2,\IR) \times SL(2,\IR)$. Therefore,
\be
\vol( O_{\IZ}(Q) \backslash O_{\IR}(Q) / O(2) \times O(2)  ) = (\sqrt{2})^4 \frac{1}{4} \vol(\CF \times \CF)= \vol(\CF\times \CF)
\ee
where the volume form on $\CF$ is induced from the metric $\Tr(A^{-1} dA)^2$ on $SL(2,\IR)$.
Using a standard KAN decomposition so that $\tau = x + \I y = A\cdot \I$ we find that
this metric is half of the standard Poincar\'e metric, and $\vol(\CF) = \pi/3$ in the
standard metric. Putting all these facts together we get
\be
\vol( O_{\IZ}(Q_{2,2}) \backslash O_{\IR}(Q_{2,2})) = \frac{4\pi^4}{9}
\ee
in agreement with \eqref{eq:Mass-Lrs}, provided we take \eqref{eq:cpt-norm}.

\section{Extremal Polar Coefficient For $S^m T4$}\label{App:T4}

We use equation (5.16) of \cite{Maldacena:1999bp} to give the formula
for the generating function of the elliptic genera of $S^m T4$:

\be\label{eq:Ell-T4}
\begin{split}
\left( \sum_{m=0}^\infty p^{m+1} {\rm Ell}(q,y; S^{m} T4)  \right) & =
\frac{\fP}{y-2 + y^{-1}}
\sum_{m\geq 1,n\geq 0, \ell\in \IZ} \frac{\hat c(nm,\ell) p^m q^n y^\ell}{(1-p^m q^n y^\ell)^2} \\
\end{split}
\ee
where
\be
\fP =
\prod_{n=1}^\infty \frac{(1-q^n)^4}{(1-y q^n)^2 (1- y^{-1} q^n)^4}
\ee
and
\be
\sum_{n\geq 0, \ell \in \IZ} \hat c(n,\ell) q^n y^{\ell} = - \frac{ (\vartheta_1(z\vert \tau))^2}{\eta^6}
= (y-2 + y^{-1}) \fP^{-1}
\ee

For the extremal polar coefficient we are interested in the coefficient of $p^m q^0 y^m$. Since we want the
$q^0$ term we can set $q=0$ and then the right hand side of \eqref{eq:Ell-T4} reduces to
\be
\sum_{m\geq 1} p^m \sum_{s\vert m} s \frac{y^s -2 + y^{-s}}{y-2 + y^{-1}}
\ee
so the $q^0$ term of ${\rm Ell}(q,y; S^{m} T4)$ is
\be
\sum_{s\vert (m+1) } s \frac{y^s -2 + y^{-s}}{y-2 + y^{-1}} = (m+1)y^m + \cdots + (m+1)y^{-m}
\ee
and hence $\fe(S^m T4) = m+1$.


\begin{thebibliography}{99}


\bibitem{Ashok:2003gk}
  S.~Ashok and M.~R.~Douglas,
  ``Counting flux vacua,''
  JHEP {\bf 0401}, 060 (2004)
  [hep-th/0307049].


\bibitem{Beauville}
A.~Beauville, ``Holomorphic symplectic geometry: a problem list,''
arXiv:1002.4321;
 “Riemannian holonomy and algebraic geometry,” math.AG/9902110;
``Vari\'et\'es k\"ahl\'eriennes dont la premi\`ere classe de Chern est nulle,'' J. of Diff.
Geometry 18, 755-782 (1983).

\bibitem{Belin:2014fna}
  A.~Belin, C.~A.~Keller and A.~Maloney,
   ``String Universality for Permutation Orbifolds,''
  Phys.\ Rev.\ D {\bf 91}, no. 10, 106005 (2015)
  [arXiv:1412.7159 [hep-th]].


\bibitem{BG}
M. Belolipetsky and W.T.~ Gan,   ``The mass of unimodular lattices,''
Journal of Number Theory, Vol. \textbf{114}(2005) p. 221.

\bibitem{Benjamin:2015hsa}
  N.~Benjamin, M.~C.~N.~Cheng, S.~Kachru, G.~W.~Moore and N.~M.~Paquette,
  ``Elliptic Genera and 3d Gravity,''
  arXiv:1503.04800 [hep-th].



\bibitem{Cassels} J.W.S. Cassels, \emph{Rational Quadratic Forms}, Dover, 1978

\bibitem{SPLG} J.H. Conway and N.J.A. Sloane, \emph{Sphere Packings, Lattices and Groups},
Springer Verlag, Second Edition

\bibitem{Dijkgraaf:1996xw}
  R.~Dijkgraaf, G.~W.~Moore, E.~P.~Verlinde and H.~L.~Verlinde,
  ``Elliptic genera of symmetric products and second quantized strings,''
  Commun.\ Math.\ Phys.\  {\bf 185}, 197 (1997)
  [hep-th/9608096].


\bibitem{Dijkgraaf:1998gf}
  R.~Dijkgraaf,
  ``Instanton strings and hyperKahler geometry,''
  Nucl.\ Phys.\ B {\bf 543} (1999) 545
  [hep-th/9810210].



\bibitem{Douglas:2005hq}
  M.~R.~Douglas and Z.~Lu,
  ``Finiteness of volume of moduli spaces,''
  hep-th/0509224.

\bibitem{Douglas:2006zj}
  M.~Douglas and Z.~Lu,
   ``On the geometry of moduli space of polarized Calabi-Yau manifolds,''
  math/0603414 [math-dg].

\bibitem{Douglas:2006es}
  M.~R.~Douglas and S.~Kachru,
  ``Flux compactification,''
  Rev.\ Mod.\ Phys.\  {\bf 79}, 733 (2007)
  [hep-th/0610102].

\bibitem{ErdosLehner} P. Erd\"os and J. Lehner, ``The distribution of the number of
summands in the partition of a positive integer,'' Duke Math. Journal
\textbf{8}(1941)335-345.

\bibitem{GHY}
W.T.~ Gan, J.P. Hanke, and J-K. Yu, ``On an exact mass formula of Shimura,''
 Duke Mathematical Journal, Vol. \textbf{107}(2001) p. 103.


\bibitem{Ginsparg:1986bx}
  P.~H.~Ginsparg,
  ``Comment on Toroidal Compactification of Heterotic Superstrings,''
  Phys.\ Rev.\ D {\bf 35}, 648 (1987).


\bibitem{Giveon:1990era}
  A.~Giveon, N.~Malkin and E.~Rabinovici,
  ``On Discrete Symmetries and Fundamental Domains of Target Space,''
  Phys.\ Lett.\ B {\bf 238}, 57 (1990).



\bibitem{Hartman}
 T.~Hartman, C.~A.~Keller and B.~Stoica,
  ``Universal Spectrum of 2d Conformal Field Theory in the Large c Limit,''
  arXiv:1405.5137 [hep-th].


\bibitem{Horne:1994mi}
  J.~H.~Horne and G.~W.~Moore,
   ``Chaotic coupling constants,''
  Nucl.\ Phys.\ B {\bf 432}, 109 (1994)
  [hep-th/9403058].

\bibitem{Keller:2011xi}
  C.~A.~Keller,
  ``Phase transitions in symmetric orbifold CFTs and universality,''
  JHEP {\bf 1103}, 114 (2011)
  [arXiv:1101.4937 [hep-th]].


\bibitem{Maldacena:1999bp}
  J.~M.~Maldacena, G.~W.~Moore and A.~Strominger,
  ``Counting BPS black holes in toroidal Type II string theory,''
  hep-th/9903163.



\bibitem{MirandaMorrison}
R. Miranda and D.R. Morrison, ``Embeddings Of Integral Quadratic Forms,''s
http://www.math.ucsb.edu/$~{}$drm/manuscripts/eiqf.pdf


\bibitem{Moore:1993zc}
  G.~W.~Moore,
   ``Finite in all directions,''
  Yale Univ. New Haven - YCTP-P12-93 (93/05,rec.Jun.) 66 p. e: LANL hep-th/9305139
  [hep-th/9305139].


\bibitem{Narain:1986am}
  K.~S.~Narain, M.~H.~Sarmadi and E.~Witten,
  ``A Note on Toroidal Compactification of Heterotic String Theory,''
  Nucl.\ Phys.\ B {\bf 279}, 369 (1987).

\bibitem{Nikulin} V.~V.~ Nikulin,
``Integral symmetric bilinear forms and some of their applications,''
Math. USSR Izvestija,  Vol. 14 (1980), No. 1, pp.103-167


\bibitem{Ranganathan:1993vj}
  K.~Ranganathan, H.~Sonoda and B.~Zwiebach,
   ``Connections on the state space over conformal field theories,''
  Nucl.\ Phys.\ B {\bf 414}, 405 (1994)
  [hep-th/9304053].


\bibitem{Seiberg:1999xz}
  N.~Seiberg and E.~Witten,
   ``The D1 / D5 system and singular CFT,''
  JHEP {\bf 9904}, 017 (1999)
  [hep-th/9903224].

\bibitem{Siegel-IndefiniteForms}
  C.L.~Siegel, ``On the theory of indefinite quadratic forms,''
  Ann. Math. Vol \textbf{45}(1944) p. 577.


\bibitem{SzalayTuran} M. Szalay and P. Tur\'an, ``On some problems of the
statistical theory of partitions with application to characters of the
symmetric group. I,'' Acta Math. Acad. Scient. Hungaricae, Vol. 29 (1977), pp. 361-379.


\bibitem{VershikYakubovich} A. Vershik and Yu. Yakubovich, ``The Limit Shape And
Fluctuations Of Random Partitions Of Naturals With Fixed Number Of Summands,''
Moscow Mathematical Journal, Vol. 1 (2001) pp.457-468

\bibitem{Zamolodchikov:1986gt}
  A.~B.~Zamolodchikov,
 ``Irreversibility of the Flux of the Renormalization Group in a 2D Field Theory,''
  JETP Lett.\  {\bf 43}, 730 (1986)
  [Pisma Zh.\ Eksp.\ Teor.\ Fiz.\  {\bf 43}, 565 (1986)].

\end{thebibliography}
\end{document}